\documentclass[12pt,preprint]{aastex}

\shortauthors{Authors}
\shorttitle{}
\begin{document} 

\newcommand\degd{\ifmmode^{\circ}\!\!\!.\,\else$^{\circ}\!\!\!.\,$\fi}
\newcommand{\etal}{{\it et al.\ }}
\newcommand{\uv}{(u,v)}
\newcommand{\rdm}{{\rm\ rad\ m^{-2}}}
\newcommand{\msuny}{{\rm\ M_{\sun}\ y^{-1}}}
\newcommand{\mylesssim}{\stackrel{\scriptstyle <}{\scriptstyle \sim}}
\newcommand{\lsim}{\stackrel{\scriptstyle <}{\scriptstyle \sim}}
\newcommand{\gsim}{\stackrel{\scriptstyle >}{\scriptstyle \sim}}
\newcommand{\sci}{Science}
\newcommand{\boo}{Bo\"{o}tes\ }
\newcommand{\Ha}{{{\rm H}\ensuremath{\alpha}}}
\newcommand{\HI}{\ion{H}{1}}

\def\kbar{{\mathchar'26\mkern-9mu k}}
\def\totd{{\mathrm{d}}}
\newcommand{\be}{\begin{equation}}
\newcommand{\ee}{\end{equation}}

\title{Comparing H$\alpha$ and \ion{H}{1} Surveys as Means to a
Complete Local Galaxy Catalog in the Advanced LIGO/Virgo Era}

\author{
Brian D.~Metzger\altaffilmark{1},
David L.~Kaplan\altaffilmark{2}, 
Edo Berger\altaffilmark{3}
}

\altaffiltext{1}{Department of Astrophysical Sciences, Peyton Hall,
Princeton University, Princeton, NJ 08542, USA;
bmetzger@astro.princeton.edu}

\altaffiltext{2}{Physics Department, University of Wisconsin -
Milwaukee, Milwaukee, WI 53211; kaplan@uwm.edu}

\altaffiltext{3}{Harvard-Smithsonian Center for Astrophysics, 60
Garden Street, Cambridge, MA 02138; eberger@cfa.harvard.edu}

\begin{abstract}

Identifying the electromagnetic counterparts of gravitational
wave (GW) sources detected by upcoming networks of advanced
ground-based interferometers will be challenging due in part to the
large number of unrelated astrophysical transients within the $\sim
10-100$ deg$^{2}$ sky localizations.  A potential way to greatly reduce the
number of such false positives is to limit detailed follow-up to only
those candidates near galaxies within the GW sensitivity range of
$\sim 200$ Mpc for binary neutron star mergers.  Such a strategy is
currently hindered by the fact that galaxy catalogs are grossly
incomplete within this volume.  Here we compare two methods for completing the local galaxy catalog: (1) a narrow-band
\Ha\ imaging survey; and (2) an \HI\ emission line radio survey.  Using \Ha\
fluxes, stellar masses ($M_{\star}$), and star formation rates (SFR)
from galaxies in the Sloan Digital Sky Survey (SDSS), combined with
\HI\ data from the GALEX Arecibo SDSS Survey and the Herschel
Reference Survey, we estimate that a \Ha\ survey with
a luminosity sensitivity of $L_{\rm H\alpha}=10^{40}$ erg s$^{-1}$ at
200 Mpc could achieve a completeness of $f_{\rm SFR}^\Ha \approx 75\%$ with respect to total SFR, but only $f_{\rm M_{\star}}^\Ha\approx 33\%$ with respect to $M_{\star}$ (due to lack of
sensitivity to early-type galaxies).  These numbers are significantly lower than those achieved by an idealized spectroscopic survey due to the loss of H$\alpha$ flux resulting from resolving out nearby galaxies and the inability to correct for the underlying stellar continuum.  An \HI\ survey with sensitivity similar to the proposed WALLABY survey on ASKAP could achieve $f_{\rm SFR}^{\rm H\,I}\approx 80\%$ and $f_{\rm M_{\star}}^{\rm H\,I}\approx 50\%$, somewhat higher than that of the H$\alpha$ survey.  Finally, both \Ha\ and \HI\ surveys should achieve $\gtrsim 50\%$ completeness with respect to the host galaxies of short duration gamma-ray bursts, which may trace the population of binary neutron star mergers.
\end{abstract}

\keywords{gamma rays: bursts$-$gravitational waves$-$galaxies:
distances and redshifts$-$radio lines: galaxies}

\section{Introduction}
\label{sec:intro}

The inspiral and coalescence of neutron star binaries (NS-NS) are the
most likely astrophysical sources for direct detection with the
upcoming advanced networks of ground-based gravitational wave (GW)
interferometers, such as Advanced LIGO and Virgo (hereafter
aLIGO/Virgo; \citealt{Abramovici+92,Caron+99,Acernese+09,Abadie+10}).
Maximizing the science achievable from such detections will require
the identification and localization of an associated electromagnetic
(EM) counterpart (e.g.,
\citealt{Bloom+09,Phinney09,Metzger&Berger12}, hereafter MB12).  The EM-GW link is
crucial for identifying the host galaxy and distance of the merger,
for placing the mergers in an astrophysical context, for studying the
hydrodynamics of matter during the merger process, and potentially for
lifting degeneracies associated with the inferred binary parameters
(e.g., \citealt{Hughes&Holz03}, \citealt{Kelley+12}).

One commonly-discussed EM counterpart of NS-NS mergers is a
short-duration gamma-ray burst (GRB); see \citet{Fong+10} and
\citet{Berger11} for evidence favoring this association.
Unfortunately, on-axis short GRBs occur at a low rate of $\lesssim 1$
yr$^{-1}$ within the $\sim 200$ Mpc range for NS-NS detections by
aLIGO/Virgo (\citealt{Nakar+06}; \citealt{Fong+10}; MB12), so the joint detection of a short GRB with a GW event is expected to be rare.  More
isotropic optical counterparts, such as an off-axis GRB afterglow
(\citealt{Coward+11}; \citealt{vanEerten&MacFadyen11}; MB12) or supernova-like emission powered by the decay of radioactive ejecta (``kilonova''; \citealt{Li&Paczynski98, Metzger+10,Roberts+11,Goriely+11,Rosswog+12}), may instead represent more promising counterparts for the bulk of GW-detected events (MB12); the typical timescale for these counterparts is a few days.  Delayed radio emission, powered by the interaction of (non-)relativistic ejecta with the surrounding environment, may also be detectable if the merger occurs in a sufficiently dense medium, with a timescale of months to decades \citep{Nakar&Piran11,Piran+12}.

Regardless of physical origin, a major challenge to identifying EM
counterparts is the expected poor sky localizations of $\sim 10-100$
deg$^{2}$ for networks of GW detectors \citep{Fairhurst09,Nissanke+11, Nissanke+12}.  In the optical band, this large
area will necessitate deep and rapid follow-up with wide-field survey
instruments.  Still, an even greater challenge may be the large number
of false positives in these wide fields (e.g., background supernovae,
M dwarf flares, shock break-out events, AGN flares; see
\citealt{Kulkarni&Kasliwal09}; MB12; \citealt{Nissanke+12}).  Given the rapid evolution of
the predicted optical signal, such false positives need to be quickly eliminated so that candidate counterparts could be followed up with
deeper photometry or spectroscopy before the transient fades.  Although rapid follow-up is not as essential at radio frequencies, false positives could also be of concern in this case  (e.g., radio supernovae, tidal disruption events, AGN flares), given the poorer angular resolution of wide-field radio survey instruments and our lack of knowledge about the transient radio sky on the timescales and depths of the expected merger counterparts (e.g., \citealt{Frail+12}).    

The contamination from false positives could be reduced by restricting
the search volume to locations near\footnotemark\footnotetext{Although
it is possible that some mergers will occur far outside of the host
galaxy of their stellar progenitors (e.g.,
\citealt{Narayan+92,Kelley+10}), even offsets as large as tens of kpc
(as characterize the observed offsets of short GRBs;
\citealt{Berger10,Fong+10}) would still reduce the required search
area substantially as compared to a search conducted with no
information on sky position.} galaxies within the sensitivity range of
aLIGO/Virgo ($z \lesssim 0.046$; e.g., \citealt{Abadie+10}).  This
approach has the potential to reduce the search area from tens of
square degrees to $\lesssim 0.5$ deg$^2$, and hence to reduce the
number of false positives by a factor of $\sim 10^3$
\citep{Kulkarni&Kasliwal09}.  However, to implement this strategy with
success, the census of local galaxies must be reasonably accurate and
complete.  Unfortunately, this is currently not the case: the galaxy
catalogs used for recent LIGO/Virgo GW follow-up (\citealt{Abadie+12}; \citealt{Evans+12}) are estimated to be only
$\approx 60\%$ complete (with respect to $B$-band luminosity) at 100
Mpc \citep{Kopparapu+08,White+11}, in which case the true counterpart
(most likely at a larger distance) could easily be missed.
Furthermore, although $B$-band completeness is a reasonable proxy for
the NS-NS merger population if the latter traces current star
formation (e.g., \citealt{Belczynski+02}), completeness with respect
to stellar mass is also relevant if a sizable fraction of mergers
occur in early-type galaxies with older stellar populations (e.g.,
\citealt{OShaughnessy+08}).

In this paper we compare two approaches for completing the local
galaxy catalog.  The first strategy is a narrow-band \Ha\ survey ($\S\ref{sec:Halpha}$), an idea that has been discussed
previously (e.g., \citealt{Rau+09}) but is fleshed out in detail here.  The other strategy, explored here for the first
time, is a wide-field \HI\ survey ($\S\ref{sec:HI}$), as is already
being planned as a main science driver for future wide-field radio
arrays.  Our main conclusion is that neither \Ha\ nor \HI\ surveys of
planned sensitivity are sufficient to fully complete the galaxy
catalog, especially with respect to stellar mass.  However, even with
respect to star formation completeness, an \Ha\ survey suffers from a
significant loss of flux (and hence sensitivity and completeness) from
spatially resolving the disks of the galaxies and from the effects of Balmer
absorption in the underlying stellar continuum.  In
$\S\ref{sec:discussion}$ we discuss the merits of combining \Ha\ and
\HI\ surveys, and explore the relevance of our results in the context
of the host galaxies of short GRBs ($\S\ref{sec:shortGRB}$).

\section{Narrow-Band \Ha\ Imaging Survey}
\label{sec:Halpha}

Most studies in the past have assumed that the rate of NS-NS mergers
in the local universe traces $B$-band luminosity
\citep{Phinney91,Kopparapu+08}, as would be expected if the rate of
mergers is proportional to the current star formation rate (SFR).
This naturally led to consideration of narrow-band \Ha\ imaging
surveys to complete the local galaxy catalog since \Ha\ emission is
closely tied to on-going star formation activity (e.g.,
\citealt{Kennicutt92,Gallagher&Gibson94,Gallego+95,Kennicutt98,Fujita+03})
and imaging surveys can efficiently identify and help characterize
galaxies in a given redshift range \citep{Dale+08,Dale+10,Ly+11}.  In
this section we address the completeness of such a survey as a
function of depth in terms of star formation and stellar mass, and
discuss the associated challenges.

We first note that while SFR is certainly an important quantity, it is
also possible that a sizable fraction of mergers instead trace total
stellar mass, as determined by the distribution of merger times and
the star formation history (\citealt{Belczynski+02,Belczynski+06}; see
Figure 11 of \citealt{OShaughnessy+08}).  \citet{Leibler&Berger10}
find evidence from the host galaxies of short GRBs that mergers may
track a combination of SFR and stellar mass, in rough analogy with
Type Ia SNe (e.g., \citealt{Scannapieco&Bildsten05}).  Given this
possibility, in what follows we explore completeness with respect to
both stellar mass and SFR independently.

In addressing the issue of \Ha\ completeness we pursue an empirical
approach, rather than directly making use of derived relations between
\Ha\ luminosity and various galaxy properties.  Our primary source of
data are galaxy catalogs with measured \Ha\ fluxes ($F_\Ha$) provided
by the MPA-JHU emission line
analysis\footnotemark\footnotetext{\url{http://www.mpa-garching.mpg.de/SDSS/DR7/}}
from the Sloan Digital Sky Survey (SDSS) Data Release 7 (e.g.,
\citealt{Strauss+02,Abazajian+09}).  Each galaxy is characterized by a
stellar mass (determined by photometric modeling;
\citealt{Kauffmann+03,Salim+07}) and SFR (determined from emission
lines; \citealt{Brinchmann+04}).  We correct the measured \Ha\ fluxes
to account for the finite angular size of the spectroscopic fiber
relative to the galaxy by scaling the fluxes by the difference of the
total $r$-band magnitude (keyword \texttt{r\_petro}) to that contained
in the spectroscopic fiber (keyword \texttt{r\_fiber}), as described
in \citet{Hopkins+03} and \citet{Brinchmann+04}.  This assumes uniform
\Ha\ surface brightness (see below for further discussion) and allows us to study the completeness achieved by an idealized survey that picks up the entire H$\alpha$ flux from the galaxy.  We also increase the quoted flux errors by a factor of 2.47 (as suggested by the MPA-JHU pipeline) and set a conservative threshold on the signal-to-noise ratio of ${\rm S/N}\gtrsim 5$ for an \Ha\ detection;
values with ${\rm S/N}\lesssim 5$ are treated as upper limits.  Finally, we remove all galaxies with ``flagged'' SFRs (keyword
\texttt{flagSFR} = 0) or redshifts (keyword \texttt{z\_warning} = 1).

Figure \ref{fig:Ha_complete} shows our results for the fraction
(``completeness'') of the total stellar mass ({\it red}) and SFR ({\it
blue}) within 200 Mpc as a function of the survey flux
depth\footnotemark\footnotetext{We do not discuss here whether the
\Ha\ survey should be framed as being defined by a flux or equivalent
width threshold (e.g., see the discussion in \citealt{sas+00}).  This
distinction will have a minor effect on our conclusions.} ($F_{\rm
lim,\Ha}$) for an idealized spectroscopic survey.  Completeness is calculated as the
fraction of the total SFR or the total stellar mass in all galaxies with $F_\Ha\gtrsim F_{\rm lim,\Ha}$, normalized to the total SFR or mass of all galaxies in the redshift range $z<0.046$.  As the flux limit
decreases below a ${\rm few}\times 10^{-15}$ erg s$^{-1}$ cm$^{-2}$ the
completeness becomes increasingly uncertain (shown by the hatched uncertainty bands), as increasing fractions of galaxies have only
upper limits on $F_\Ha$.  The miminum (maximum) completeness in this
case is calculated by assuming that none (all) would be detections at
$F_\Ha<F_{\rm lim,\Ha}$.

An idealized survey to a depth $F_{\rm lim,\Ha} = 2\times 10^{-15}$
erg s$^{-1}$ cm$^{-2}$ (a luminosity of $L_{\rm lim,\Ha} = 10^{40}$
erg s$^{-1}$ at 200 Mpc), as could be reached with a few minute
integration per pointing on a meter-class telescope such as the
Palomar Transient Factory \citep{Law+09}, could achieve a completeness of $f_{\rm SFR}^\Ha\approx 97\%$
with respect to SFR, but only $f_{\rm M\star}^\Ha \approx 60-80\%$
with respect to total stellar mass.  It is not surprising that $f_{\rm
SFR}^\Ha>f_{\rm M_{\star}}^\Ha$ given the known correlation between
\Ha\ luminosity and SFR \citep{Kennicutt98,Kewley+02}.  Lower
completeness with respect to stellar mass is consistent with the fact
that $\approx 60\%$ of mass in the local universe is in early-type galaxies \citep{Bell+03}, while they account for only $\approx 10$\% of the \Ha\
luminosity density at $z \lesssim 0.046$ (e.g.~\citealt{Nakamura+03}).

For comparison in Figure~\ref{fig:Ha_complete} we also plot the
completeness with respect to $B$-Band luminosity using data from the
11HUGS survey of galaxies at $<11$ Mpc \citep{Lee+07,Kennicutt+08}, and
with respect to total \Ha\ luminosity (using the $z = 0$ \Ha\
luminosity function from \citealt{Gallego+95}; cf.~\citealt{Ly+07};
\citealt{Dale+10}).  $B$-Band completeness ($f_{B}^\Ha$) tracks closer
to SFR than to stellar mass, with $f_{ B}^\Ha\approx 87\%$
completeness at $L_{\rm \Ha}=10^{40}$ erg s$^{-1}$.  This is not
unexpected, as blue light tracks ongoing star formation, although not
as tightly as \Ha\ luminosity.  

The estimates of mass and SFR completeness in
Figure~\ref{fig:Ha_complete} are only appropriate for an idealized
survey, such as that realized through direct spectroscopy of the
target galaxies (as in SDSS), but with infinite aperture.  For an actual
imaging surveys, absorption in the underlying stellar continuum can
reduce the perceived \Ha\ flux when integrated over a filter (e.g.,
\citealt{Meurer+06}).  While for young star-forming galaxies this will
generally be a minor correction \citep{Brinchmann+04,Meurer+06}, the
correction can be significant for early-type galaxies.  In
Figure~\ref{fig:3spectra} we show a comparison between the SDSS
spectra of three galaxies with similar \Ha\ fluxes, but where the
integrated flux in a narrow-band H$\alpha$ filter is reduced by
increasing amounts of Balmer absorption from the stellar continuum.
In the bottom spectrum, the \Ha\ equivalent width (EW) is much greater
than the correction due to stellar absorption (EW$_{\star}$), in which
case the galaxy would be detected regardless of the effects of stellar
absorption.  However, in the top spectrum EW and EW$_{\star}$ are
comparable, and the galaxy might escape detection in a narrow-band
imaging survey.

To explore what completeness could be achieved by an actual imaging
survey that cannot correct the \Ha\ fluxes for stellar absorption, we
recalculated the minimum mass and SFR completeness using the fluxes
from SDSS which are instead calculated from the line equivalent widths
computed via integration over broad wavelength ranges.  The results are shown
with dashed lines in Figure \ref{fig:Ha_complete}.  Although the
effect on SFR completeness $f_{\rm SFR}^\Ha$ is relatively minor (less
than a few percent, and comparable to the corrections considered in
\citealt{Meurer+06}), the completeness with respect to stellar mass
decreases to $f_{\rm M_{\star}}^\Ha\approx 47\%$.  This difference
arises because Balmer absorption is strong and \Ha\ is weak in
galaxies with relatively old stellar populations and low levels of
star formation activity, i.e., those that contribute to stellar mass
but not SFR completeness.

Another challenge of a realistic survey is that nearby galaxies
generally have angular sizes larger than the typical
point-spread-function of a ground-based imaging survey.  This effect
can reduce the effective survey sensitivity compared to that for point
sources (i.e., the survey is actually defined by a limiting surface
brightness instead of a limiting flux).  The actual correction between
limiting surface brightness and limiting flux depends on the
background noise level, the typical seeing, and the distribution of
galaxy sizes.  In Figure~\ref{fig:radii} we show the distribution of
effective angular radii ($\mathcal{R}_{\rm eff}\equiv \mathcal{R}_{\rm
ap,SDSS} \times 10^{(\texttt{r\_fiber}-\texttt{r\_petro})/5}$) of all galaxies in our sample, where $\mathcal{R}_{\rm ap,SDSS}$ is the $1.5\arcsec$ radius spectral aperture of SDSS.  A significant number of sources have radii of $\gtrsim 5\arcsec$, meaning that they will cover $\gtrsim 10$ seeing disks for typical conditions, although this will be somewhat mitigated by the fact that \Ha\ typically comes from
localized \ion{H}{2} regions which have higher-than-average surface
brightnesses.

We include the effects of Balmer absorption and the galaxy size
distribution to determine more realistic completeness fractions.  We use the uncorrected SDSS \Ha\ fluxes (i.e., not scaling the
\Ha\ fluxes by the total $r$-band magnitude, as described above).
This gives an effective seeing disk of $1.5\arcsec$, comparable to typical ground-based observing
conditions, and basically assumes that the H$\alpha$ flux outside the
central portion is below the detectability threshold.  We recompute
the stellar mass and SFR completeness using this pessimistic (``imaging'') scenario, and the resulting minimum completeness are shown with solid lines in
Figure~\ref{fig:Ha_complete}.  For a (point source) luminosity $L_{\rm
H\alpha} = 10^{40}$ erg s$^{-1}$, the SFR completeness is reduced to
$f_{\rm SFR}^\Ha \approx 76\%$, while the stellar mass completeness is
$f_{\rm M_{\star}}^\Ha\approx 33\%$.

In Table~\ref{table:surveycompare} we summarize the relevant
completeness values.  We give three values for the \Ha\ survey, as
discussed above.  The first corresponds to our ideal survey, where
there is no Balmer absorption and all of the flux from extended
sources is recovered.  The other two (realistic) sets include Balmer
absorption, with the optimistic version still assuming that all of the
flux from extended sources is recovered, and the pessimistic version
assuming that none of the flux from extended sources is recovered; the
true value likely is between these last two versions, i.e.~$f_{\rm SFR}^\Ha \approx 75-95\%$ and $f_{\rm M_{\star}}^\Ha \approx 30-45\%$

\section{\HI\ Survey}
\label{sec:HI}

Another strategy to complete the local galaxy catalog is via
a wide-field \HI\ (21 cm) emission line survey, looking for neutral
rather than ionized gas.  We are motivated in particular by the
planned Widefield ASKAP $L$-band Legacy All-Sky Blind surveY (WALLABY;
e.g., \citealt{Duffy+12b}), which plans to observe $\approx 75\%$ of
the sky\footnotemark\footnotetext{WALLABY may be complemented by
similar surveys in the north, such as the proposed Westerbork Northern
Sky \HI\ Survey; see
{\url{http://www.astron.nl/{\textasciitilde}jozsa/wnshs/survey\_layout.html}}.}
(declination of $-90^{\circ}$ to $+30^{\circ}$) over a timescale of
$\sim 1$ year and detect about a half-million galaxies to $z\approx
0.26$ with an estimated rms sensitivity (over a 100 kHz bandwidth) of
$\approx 0.7$ mJy.  Below we examine to what extent WALLABY will be
effective at completing the local galaxy catalog.  Throughout we
assume that \HI\ luminosity ($L_{\rm H\,I}$) scales with gas mass
($M_{\rm H\,I}$) as $L_{\rm H\,I}={2}\times 10^{25}(M_{\rm
H\,I}/M_\odot)\,{\rm erg\,s}^{-1}$ \citep{Duffy+12b}, neglecting \HI\
self-absorption (a good approximation;
\citealt{Zwaan+97}). Determining the threshold for a given survey also
requires assumptions about the velocity widths and inclinations of the
galaxies.  For our purposes we assume the limiting flux density and linewidth above, which are sufficient to detect $M_{\rm H\,I}\approx 10^{9}M_{\odot}$ at a distance of 200 Mpc
at $5\sigma$ significance \citep{Duffy+12b,Duffy+12a}.

As in the case of \Ha, we opt for an empirical approach to estimate
the achievable completeness with respect to stellar mass and SFR.  We
are limited by the absence of a single large sample of galaxies with
measured \HI\ masses (or upper limits), stellar masses, and SFRs, yet
which is unbiased with respect to galaxy population.  For this reason,
we use several different \HI\ catalogs to explore completeness with
respect to stellar mass and SFR (see Figure~\ref{fig:Mstar_HI} for the
\HI\ masses and stellar masses).

To study completeness among star forming galaxies, we use \HI\ masses
from the Arecibo Legacy Fast ALFA Survey (ALFALFA;
\citealt{Giovanelli+05}) with SDSS counterparts (for which stellar
mass and SFR are available as described in $\S\ref{sec:Halpha}$) in
the redshift range $z < 0.046$.  Although the entire
ALFALFA sample is large (about 9000 galaxies) and covers a wide range
in stellar mass ($M_{\star}\approx 10^{7.5}-10^{11.5}$ M$_{\odot}$) it
is biased towards star forming galaxies and contains only SDSS
counterparts with \HI\ detections.  A complementary data set is the
GALEX Arecibo SDSS Survey (GASS; \citealt{Catinella+10,Catinella+12}),
which is much smaller ($\approx 200$ galaxies following our cuts) but also contains deep \HI\ upper limits (down to HI masses of a few percent of the stellar mass).  Unfortunately,
even GASS does not represent a fair sample of the entire galaxy
population since it is restricted to massive galaxies
($M_{\star}\gtrsim 10^{10}$ M$_{\odot}$), which for example only
account for about half of the short GRB hosts
\citep{Leibler&Berger10}; see Figure~\ref{fig:Mstar_HI}.  For this
reason, we also use \HI\ masses and upper limits from the Herschel
Reference Survey (HRS; \citealt{Cortese+12}), a volume limited sample at $<20$ Mpc \citep{Boselli+10} that covers a relatively
broad range in stellar masses ($M_{\star}\approx 10^{8.5}-10^{11.5}$
M$_{\odot}$), covering most of the short GRB host galaxies.

In Figure~\ref{fig:HI_complete} we show our results for the
completeness achievable by an all-sky \HI\ survey with respect to
total SFR $f_{\rm SFR}^{\rm H\,I}$ ({\it blue}) and stellar mass
$f_{\rm M_{\star}}^{\rm H\,I}$ ({\it red}), as a function of survey
limiting flux ($F_{\rm lim,H\,I}$), calculated using the GASS sample
(also see Table~\ref{table:surveycompare}).  For mass completeness, we
also show the HRS sample ({\it orange}).  Note that as in the case of
H$\alpha$, completeness is uncertain at low fluxes as the result of
\HI\ upper limits, as indicated by the hatched regions.

We find that a survey similar to WALLABY could achieve an SFR completeness of $f_{\rm SFR}^{\rm H\,I}\approx 93-98\%$, but only $f_{\rm M_{\star}}^{\rm H\,I}\approx 44-50\%$ with respect to total stellar mass.  For mass completeness we quote numbers from HRS instead of GASS, since HRS samples a wider range of galaxy masses (Fig.~\ref{fig:Mstar_HI}).  Again, a high SFR completeness is expected given the
correlation between cold gas mass and SFR \citep{Bigiel+11}.  Lower
completeness with respect to stellar mass is expected since early-type
galaxies are typically deficient in \HI\ \citep{Sage+07,Oosterloo+10}
yet they contain $\approx 60\%$ of mass in the local universe
\citep{Bell+03}.  For comparison, we also shows completeness with
respect to total \HI\ mass (as determined from the local \HI\
luminosity function of \citealt{Zwaan+05}), which is found to lie
between that of stellar mass and SFR (Fig.~\ref{fig:HI_complete}; Table~\ref{table:surveycompare}).

As we discussed in the context of \Ha\ surveys, for \HI\ surveys we
should also consider the possible detrimental effect of spatially
resolving the galaxies.  However, the poorer angular resolution of ASKAP ($10-30\arcsec$) makes this less of an issue than for an
optical survey, and the expected effect on detectability is
anticipated to be minimal (see Figure 5 of \citealt{Duffy+12b}).  The
limited angular resolution of the \HI\ surveys will not affect the
follow-up of GW triggers, as a $\pm 30\arcsec$ position would be quite
sufficient and the \HI\ centroid will be known to even better
precision.  Ideally, candidate galaxies detected by WALLABY will also
be followed up to obtain optical counterparts by wide-field survey
instruments such as VISTA Hemispheric Survey \citep{Arnaboldi+07}, VST
ATLAS \citep{Capaccioli+05}, or SkyMapper \citep{Keller+07}.

\section{Discussion}
\label{sec:discussion}

Table~\ref{table:surveycompare} compares the minimum completeness to
SFR and stellar mass achieved by our fiducial \Ha\ and \HI\ surveys
based on Figures~\ref{fig:Ha_complete} and \ref{fig:HI_complete}.
Overall we find that when considering realistic surveys including the
effects of Balmer absorption and finite source sizes, an \HI\ survey
can achieve a somewhat better completeness in both SFR and $M_{\star}$ compared
to \Ha.

It is clear from our results that neither the proposed \Ha\ or \HI\
surveys will produce a local galaxy catalog that is entirely complete
with respect to stellar mass out to 200 Mpc.  Fundamentally, both \Ha\
and \HI\ trace gas, while $\approx 60\%$ of stellar mass in the local universe
is in early-type galaxies \citep{Bell+03} with little gas.  However, even completeness
with respect to SFR is unlikely to reach $100\%$ from either survey
due to the issues discussed in $\S\ref{sec:Halpha}$ and
$\S\ref{sec:HI}$.  

A remaining question is whether there are any advantages to combining
the results of separate \Ha\ and \HI\ surveys.  In Figure~\ref{fig:Halpha_HI} we show $L_{\rm \Ha}$ versus $M_{\rm H\,I}$
from the ALFALFA-SDSS sample of star forming galaxies.  We note that
an obvious correlation exists between $M_{\rm H\,I}$ and $L_{\rm
H\alpha}$ since both trace star formation, but there is also a large
scatter ($\approx 1$ dex) in the correlation where the sensitivity
threshold of WALLABY ($M_{\rm H\,I}\approx 10^{9}$ M$_\odot$ at 200
Mpc) intersects that of our fiducial \Ha\ survey ($L_{\rm H\alpha}
\approx 10^{40}$ erg s$^{-1}$ at 200 Mpc).  This scatter implies that some fraction of galaxies which are just below the detection threshold in \HI\ will be detectable in \Ha\ and vice versa.  Therefore the SFR completeness attained by combining the results of both \Ha\ and \HI\ surveys together should be more than that of either individually.  

If one were to combine \HI\ and H$\alpha$ surveys, which individually
achieve a similar completeness (as is the case for WALLABY and our
`ideal' spectroscopic H$\alpha$ survey)\footnote{The fact that a
  smaller fraction of sources lie in the top-left quadrant (detected
  in H$\alpha$ but not \HI) of Figure \ref{fig:Halpha_HI} than in the
  bottom-right (detected in \HI\ but not H$\alpha$) appears at odds
  with the comparable completeness attained by our fiducial \HI\ and
  spectroscopic \Ha\ surveys (cf.~Fig.~\ref{fig:Ha_complete},
  \ref{fig:HI_complete}); however, the ALFALFA sample includes only
  \HI\ detections and hence overestimates the fraction of \HI-detected
  galaxies.}, then from the SDSS-ALFALFA sample we estimate that
$f_{\rm SFR}$ could be increased from $\sim 95\%$ to $\sim 98\%$.
Mass completeness could be increased by a somewhat greater amount
(changing by up to $\approx 10-20\%$), however $f_{\rm M_{\star}}$ is ultimately limited to a value $\lesssim 60\%$ due to the fraction of early-type galaxies which possess neither detectable \HI\ or H$\alpha$ emission.  We conclude that combining surveys provides the greatest benefit if both \HI\ and H$\alpha$ surveys separately achieve a similar completeness, however even in this case the gains will be relatively modest.

\subsection{Connection to Short GRB Hosts}
\label{sec:shortGRB}

So far we have focused on completeness with respect to SFR and stellar mass separately.  However, it is also of interest to investigate how complete the \Ha\ or \HI\ surveys will be with respect to the known population of short GRB host galaxies.  We attempt to answer this question empirically by investigating the fraction of galaxies with properties (SFR or stellar mass $M_{\star}$) similar to those of the short GRB hosts that would be detected in \Ha\ or \HI.  

In our calculation we use short GRB host galaxy masses from \citet{Leibler&Berger10} and SFRs (or upper limits) from \citet{Berger09}, resulting in a sample of 11 galaxies.  We define the `short GRB host completeness' as the fraction of galaxies in our sample that have masses (and in some cases SFRs) within a factor of $\pm$ 0.5 dex of those of each short GRB host and that are detectable at a given \Ha\ or \HI\ flux, the result of which is then averaged over the short GRB hosts.  In the case of \Ha, the galaxy masses, SFRs, and \Ha\ fluxes are again taken from the SDSS sample ($\S\ref{sec:Halpha}$), the latter of which are not corrected for stellar absorption or galaxy size (as would characterize a purely imaging survey).  In the case of \HI\, we use galaxy masses from the HRS sample, but we cannot make a cut on the SFR since these data are not available.

Figure \ref{fig:shortGRBcomplete} shows our results for completeness as a function of \Ha\ or \HI\ flux, normalized to the sensitivity of our fiducial surveys.  In the case of \Ha, results are shown for two cases: (1) in which the galaxy samples are chosen to match both the masses and SFRs of the short GRB hosts, and (2) in a case for which the samples are chosen based just on sharing similar stellar masses with the short GRB hosts.  By this criteria, we find that our fiducial \Ha\ imaging survey could achieve a short GRB host completeness of $f_{\rm SGRB}^{\Ha} \approx 50\%$ and $\approx 25\%$ in cases (1) and (2), respectively.  Case (1) is more realistic and results in a higher completeness because most of the short GRB hosts (9 of 11) are star-forming galaxies, so the SFR cut preferentially picks out H$\alpha$-luminous galaxies.  In both cases the minimum completeness asymptotes to $\approx 80\%$ at low fluxes since 2 of 11 of the short GRB hosts are early-type galaxies, for which there are only upper limits on their SFRs and H$\alpha$ luminosities.

In the case of \HI, we find that our fiducial survey similar to WALLABY could achieve a completeness of $f_{\rm SGRB}^{\rm H\,I} \approx 45\%$.  However, this probably underestimates the true completeness for the same reason as with \Ha: most of short GRB hosts are star-forming and there is a correlation between SFR and \HI\ mass (\citealt{Bigiel+11}), yet no cut was made on SFR.  

We conclude that both \Ha\ and \HI\ surveys could achieve at least
$\sim 50\%$ completion with respect to GRB host galaxies, although
this number is likely to be substantially higher in the case of an
\HI\ survey.  There is room to increase the completeness by performing a deeper search than our fiducial survey, potentially approaching the $\sim 80\%$ maximum completeness expected if one could detect all 9 of 11 star-forming short GRB hosts at $\sim$ 200 Mpc.

\section{Conclusions}
\label{sec:conclusions}

\begin{figure}
\includegraphics[width=\textwidth]{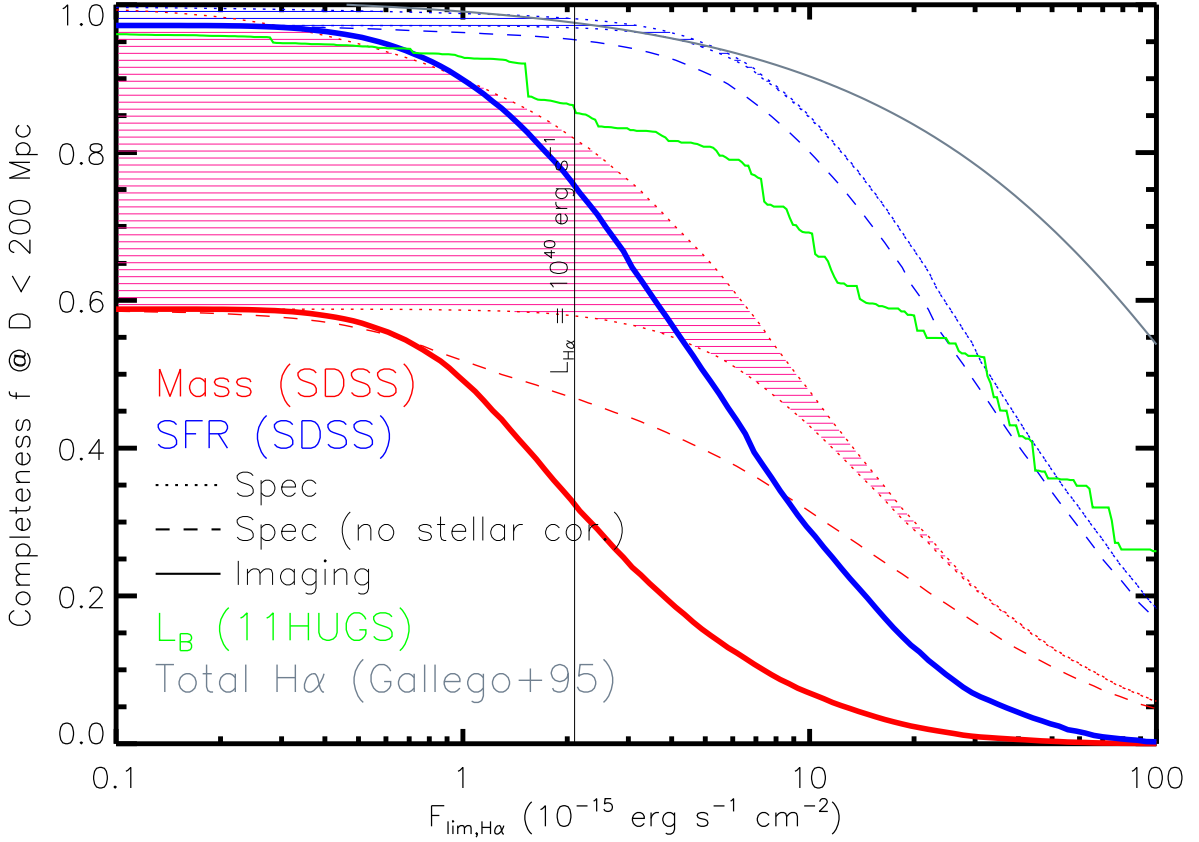}
\caption{Estimated completeness of an all-sky, narrow-band \Ha\ survey
with respect to total stellar mass ({\it red}) and total star
formation ({\it blue}) as a function of survey depth, $F_{\rm
lim,\Ha}$, calculated using \Ha\ fluxes and galaxy properties derived
from SDSS (see text).  Dotted lines show the completeness of an idealized spectroscopic survey which measures the entire H$\alpha$ luminosity of the galaxy (infinite aperture) and corrects the H$\alpha$ flux for stellar absorption; the cross-hatched region represents the uncertainties due to \Ha\ upper limits.  Dashed lines show how the minimum completeness decreases when one does not correct \Ha\ fluxes for the underlying stellar Balmer continuum, as appropriate for narrow-band imaging.  Solid lines show the minimum completeness when the \Ha\ fluxes are also not corrected for the finite angular size of the galaxy (assuming a $1.5\arcsec$ radius aperture).  These last two cases likely bracket the completeness provided by a purely imaging survey.  Also shown for comparison is completeness with respect to $B$-band luminosity ({\it green}) of the local ($<11\,$Mpc)
11HUGS survey (\citealt{Kennicutt+08}) and with respect to total \Ha\
luminosity ({\it gray}; using the \citealt{Gallego+95} luminosity
function).}
\label{fig:Ha_complete}
\end{figure}

\begin{figure} 
\includegraphics[width=\textwidth]{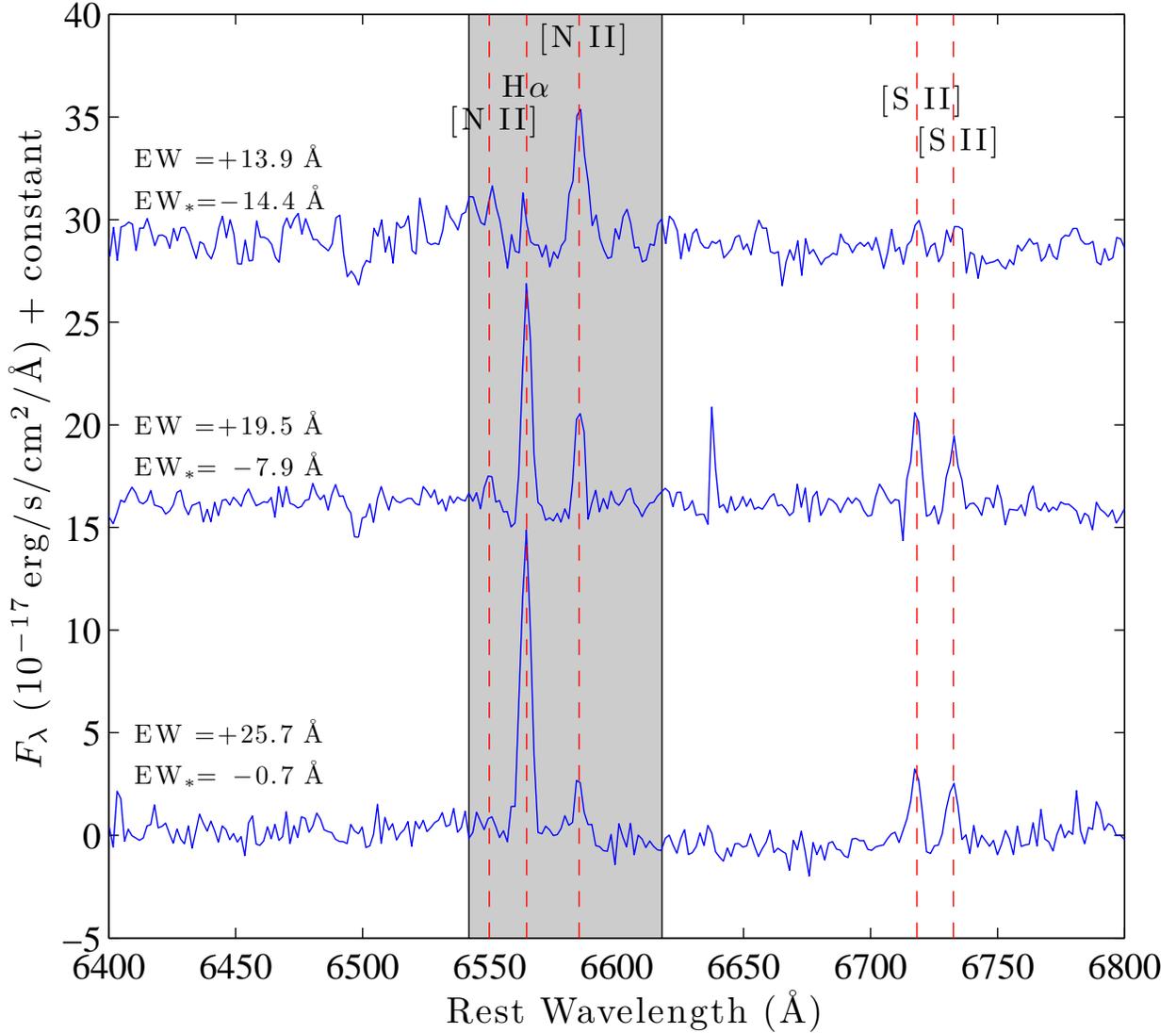}
\caption{Spectra of three galaxies from our SDSS sample which
illustrate, from bottom to top, the effect of increasing stellar
absorption on the measured H$\alpha$ equivalent width (EW) in a
narrow-band survey.  The line centers of the H$\alpha$, [\ion{N}{2}],
and [\ion{S}{2}] lines are shown with red vertical dashed lines
(wavelengths are shifted to the rest frame).  The gray shaded region
represents the approximate bandwith of a $z=0-0.01$ \Ha\ filter.  For
each galaxy we give the EW of the \Ha\ emission line and the EW of the
Balmer absorption (EW$_*$), with positive values indicating emission;
all values have been scaled to have the same continuum.  The \Ha\
fluxes (from the MPA-JHU emission line analysis) are $72\times 10^{-17}\,{\rm
erg\,s}^{-1}\,{\rm cm}^{-2}$, $58\times 10^{-17}\,{\rm erg\,s}^{-1}\,{\rm cm}^{-2}$,
and $37\times 10^{-17}\,{\rm erg\,s}^{-1}\,{\rm cm}^{-2}$, from bottom to top. It is
clear that the \Ha\ EW in the bottom galaxy greatly exceeds the Balmer
absorption (by a factor of $>30$), while in the upper galaxies the
Balmer absorption is 30\% and 100\% of the \Ha\ emission, even as the
line flux changes by only a factor of 2.  This would make detecting
the top galaxy in a narrow-band survey difficult, although the
[\ion{N}{2}] doublet will generally provide a minimum flux of $\approx
10\%$ of the original \Ha\ flux (dependent on metallicity, among other
factors; \citealt{pp04}).}
\label{fig:3spectra}
\end{figure}

\begin{figure}
\includegraphics[width=\textwidth]{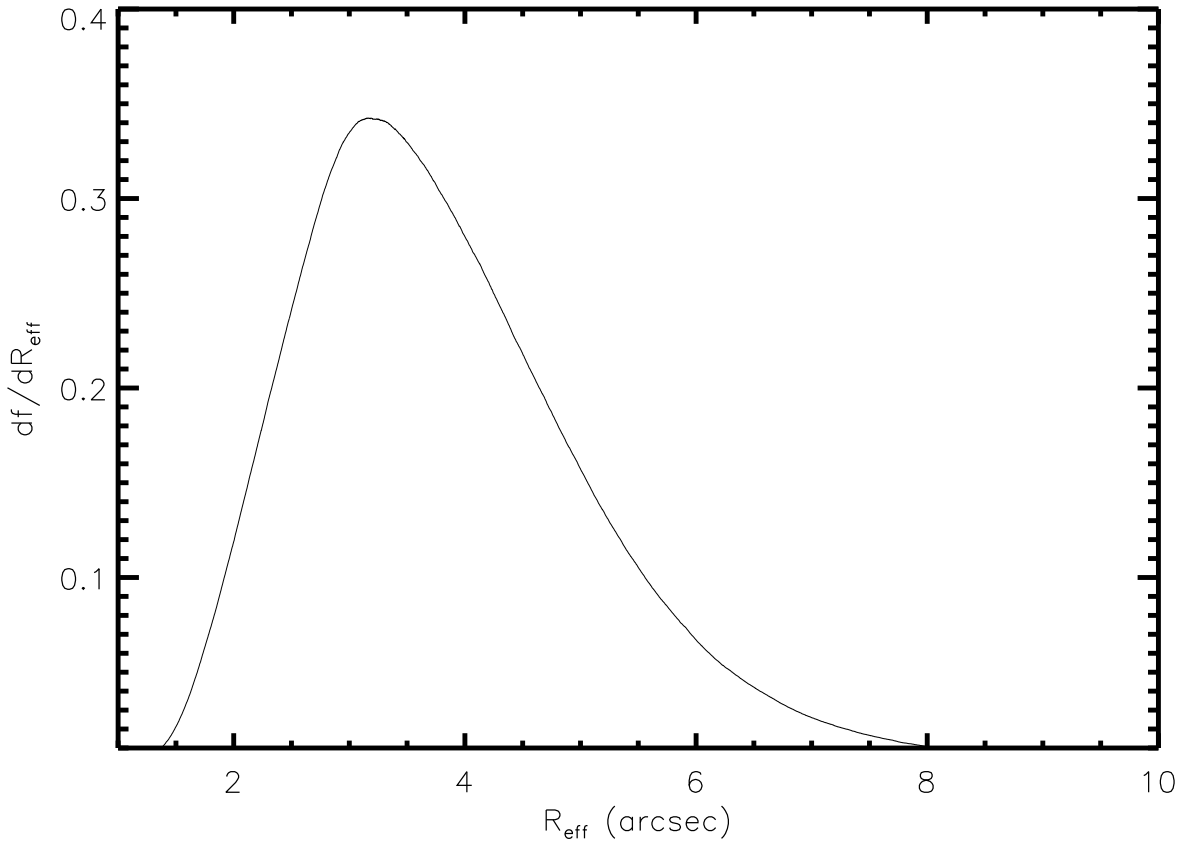}
\caption{Normalized distribution df/d$\mathcal{R}_{\rm eff}$ of
effective angular radii $\mathcal{R}_{\rm eff}$ of galaxies in our
SDSS sample at redshift $z < 0.046$ (containing 103,025
sources).  The effective radius is defined as $\mathcal{R}_{\rm eff}
\equiv \mathcal{R}_{\rm ap,SDSS}\times 10^{(\texttt{r\_fiber}-\texttt{r\_petro})/5}$,
where \texttt{r\_fiber} and \texttt{r\_petro} are the fiber and Petrosian $r$-band
magnitudes and $\mathcal{R}_{\rm ap,SDSS}$ is the $1.5\arcsec$ spectral
aperture of SDSS.  To the extent that H$\alpha$ and $r$-band have the
same surface brightness distribution, the flux sensitivity to an
extended H$\alpha$ source is reduced by a factor $\sim
(\mathcal{R}_{\rm eff}/\mathcal{R}_{\rm app})^{2}$ compared to a point
source.}
\label{fig:radii}
\end{figure}

\begin{figure}
\includegraphics[width=\textwidth]{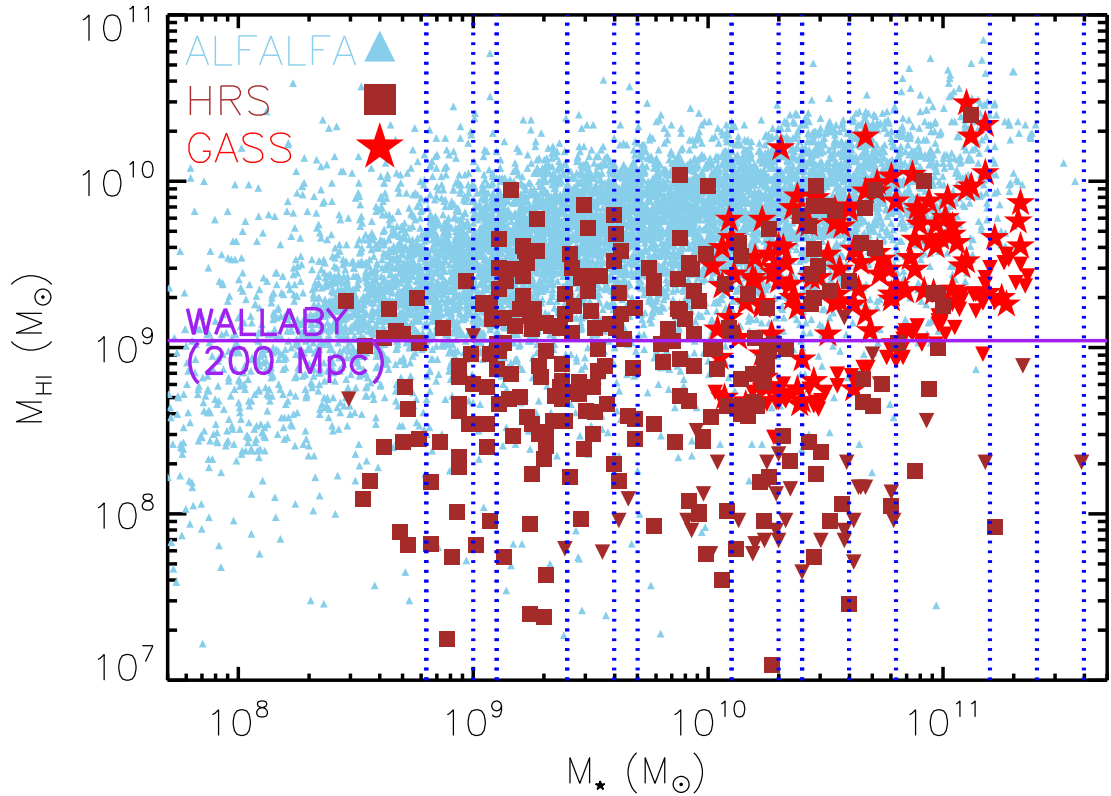}
\caption{Stellar mass $M_{\star}$ versus \HI\ mass $M_{\rm H\,I}$ from
samples of galaxies used in estimating \HI\ completeness: ALFALFA
({\it light blue points}, detections only); HRS ({\it brown}, both
detections [\textit{squares}] and upper
limits); and GASS ({\it red}, both detections [\textit{stars}] and upper limits); see text
for references.  Vertical blue lines show the masses of short GRB host
galaxies from \citet{Leibler&Berger10}, while the horizontal purple
line shows the approximate \HI\ sensitivity threshold of WALLABY at
200\,Mpc.}
\label{fig:Mstar_HI}
\end{figure}

\begin{figure}
\includegraphics[width=\textwidth]{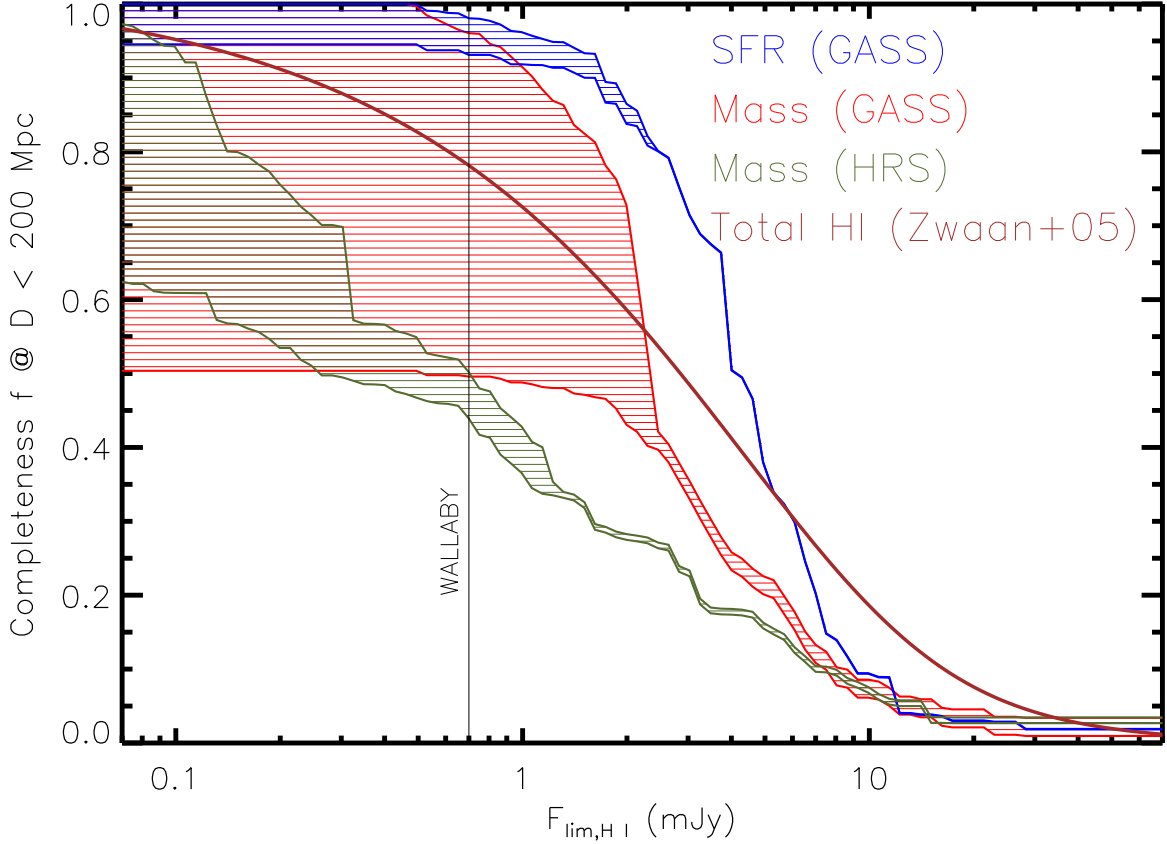}
\caption{Estimated completeness of an all-sky \HI\ survey with respect
to total stellar mass ({\it red}) and total star formation ({\it
blue}) as a function of survey depth ($F_{\rm lim,H\,I}$), calculated
using \HI\ fluxes from the GASS sample and galaxy properties derived
from the SDSS catalog (see Fig.~\ref{fig:Mstar_HI} text).  Survey depth is calculated as the 5$\sigma$ rms sensitivity over an assumed 100 kHz bandpass (see text).  We also
show the mass completeness calculated using \HI\ masses from the
Herschel Reference Survey (HRS; {\it orange}) and the completeness
with respect to total \HI\ mass ({\it brown}) calculated using the
local \HI\ luminosity function from \citet{Zwaan+05}.  The hatched regions indicate the range of uncertainty in completeness at low flux due to \HI\ upper limits.}
\label{fig:HI_complete}
\end{figure}

\begin{figure}
\includegraphics[width=\textwidth]{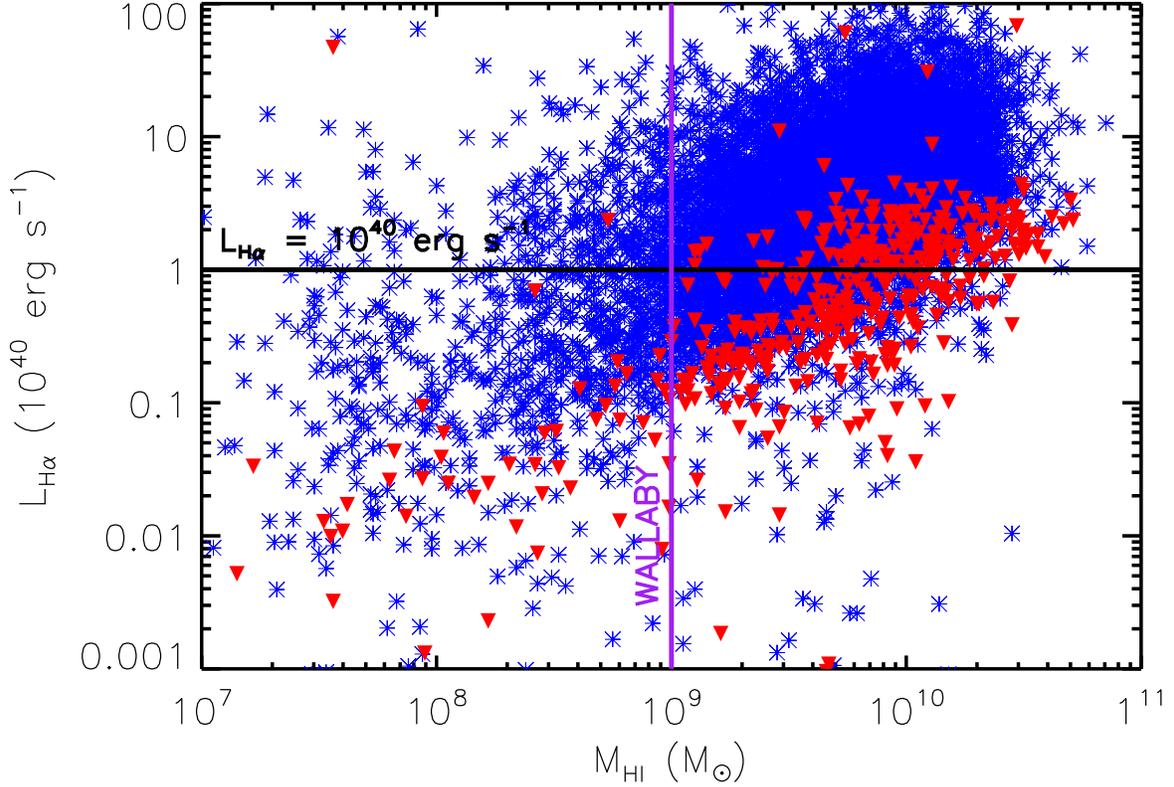}
\caption{\Ha\ luminosities $L_{\Ha}$ versus \HI\ masses $M_{\rm H\,I}$
from the ALFALFA-SDSS sample of star-forming galaxies, showing both
\Ha\ detections ({\it blue stars}) and upper limits ({\it red
triangles}).  Although a clear correlation exists between $L_{\Ha}$
and M$_{\rm H\,I}$ (since both trace star formation), the scatter in
$L_{\rm \Ha}$ is significant ($\approx 1\,$dex) near the sensitivity
threshold of an \HI\ survey similar to WALLABY ({\it purple line}).
The SFR completeness obtained by combining \Ha\ and \HI\ surveys may
thus be increased somewhat compared to that obtained by either
individually.}
\label{fig:Halpha_HI}
\end{figure}

\begin{figure}
\includegraphics[width=\textwidth]{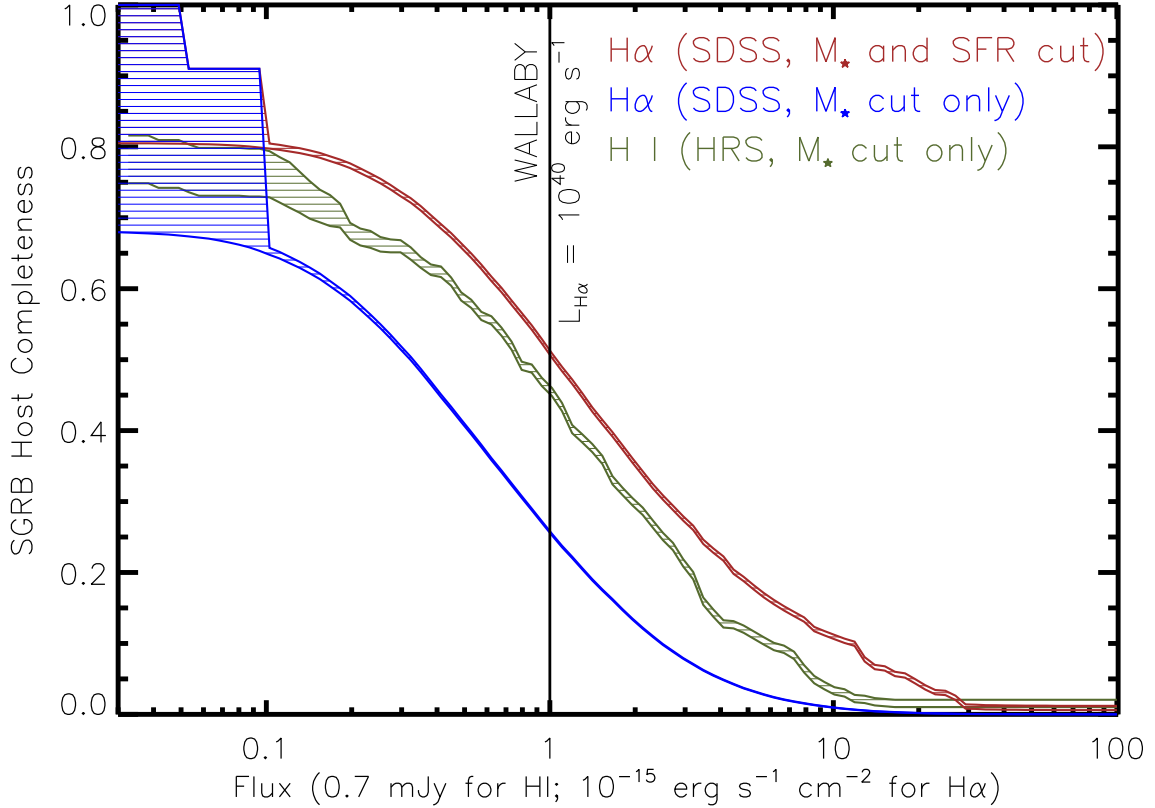}
\caption{Completeness of \Ha\ and HI surveys with respect to galaxies with properties (SFR and $M_{\star}$) similar to those of the host galaxies of short GRBs, normalized to the sensitivity of our fiducial \Ha\ imaging survey and to that of WALLABY.  In the case of \Ha, the SDSS galaxy sample is used; results are shown both using a subsample of galaxies selected based on similar stellar masses and SFRs to the short GRB hosts ({\it brown}), as well as subsamples chosen based just on similar stellar masses ({\it blue}).  In the case of \HI, the HRS sample is used and results are shown just for the subsample with similar stellar masses ({\it orange}); since most SGRB hosts are star forming, the completeness achievable by \HI\ is probably underestimated by this figure (see text).  
}
\label{fig:shortGRBcomplete}
\end{figure}

We used observed samples of nearby galaxies, drawn from SDSS
and augmented by \HI\ data from several surveys, to estimate the
completeness of fiducial \Ha\ and \HI\ surveys to SFR and stellar mass
out to a distance of 200 Mpc.  We conclude that neither \HI\ nor \Ha\
surveys to proposed depths will be entirely effective at completing
the local galaxy catalog with respect to stellar mass within 200 Mpc,
but that we can expect reasonable completenesses of $f_{\rm SFR}^\Ha
\approx 76\%$, $f_{\rm M_{\star}}^\Ha\approx 33\%$, $f_{\rm SFR}^{{\rm
H\,I}}\approx 93\%$, and $f_{\rm M_\star}^{{\rm H\,I}}\approx 44\%$.
At this point neither the large-scale \Ha\ nor \HI\ surveys discussed
here are underway, and it may be possible to alter the strategy (i.e.,
survey depth) to increase the completeness.  For instance, halving the
survey threshold of the fiducial H$\alpha$ imaging survey results in a substantial increase in the completenesses to $f_{\rm SFR}^\Ha \approx 90\%$, $f_{\rm M_{\star}}^\Ha\approx 49\%$, but for \HI\ only results in a modest increase to $f_{\rm SFR}^{{\rm
H\,I}}\approx 95\%$, and $f_{\rm M_\star}^{{\rm H\,I}}\approx 50\%$.  For \Ha, these increases are comparable to that achieved by pursuing a spectroscopic survey (Fig.~\ref{fig:Ha_complete}).  Such a deeper survey would also increase the completeness with respect to short GRB host galaxies by $\gtrsim 10\%$ (Fig.~\ref{fig:shortGRBcomplete}).  

If the survey thresholds remain as discussed here, assessing whether
or not this is a concern for aLIGO/Virgo follow-up largely depends on
the fraction of mergers that occur in early-type galaxies.  If the
host galaxies of short GRBs indeed represent a faithful sampling of
the merger population, then the current census of about a 5 to 1 ratio
of star forming to elliptical hosts \citep{Berger11} would suggest
that incompleteness of early-type galaxies is not a major concern.  On
the other hand, it is also possible that short GRBs do not represent
all mergers, or that the short GRB host population is biased against
early-type galaxies.  Current population synthesis models do allow for
a sizable fraction $\gtrsim 20-50\%$ of elliptical hosts at $z=0$
\citep{OShaughnessy+08}.

We note that our treatment of the surveys used simply-defined
empirical data sets, and some secondary effects may change the results
slightly.  For instance, neither the \Ha\ flux nor the Balmer
absorption is expected to be distributed uniformly within the
galaxies.  While we do not think that effects such as these will
greatly change our conclusions, future studies could be done with more
carefully constructed samples or could make use of observed
semi-empirical correlations between galaxy properties to directly
calculate the completeness.

\acknowledgements We thank Jarle Brinchmann and Christy Tremonti for
helpful conversations and guidance using the MPA-SDSS galaxy data;
Barbara Cantinella and Luca Cortese for helpful conversations and
guidance using the GASS and HRS \HI\ data; and Dawn Erb, Wen-fai
Fong, Alicia Soderberg, and Mansi Kasliwal for helpful discussions.  We are grateful to the KITP in Santa
Barbara for hosting the program ``Chirps, Mergers and Explosions'',
where this work began.  BDM was supported in part by NASA through Einstein Postdoctoral Fellowship grant number PF-00065.  BDM also acknowledges support from the Lyman Spitzer, Jr. Fellowship awarded by the Department of Astrophysical Sciences at Princeton University.  This research was also supported in part by the
National Science Foundation under Grant Nos.\ PHY11-25915 (KITP),
AST-1008353 (DLK), and AST-1107973 (EB).  Funding for the SDSS and
SDSS-II has been provided by the Alfred P.\ Sloan Foundation, the
Participating Institutions, the National Science Foundation, the U.S.\
Department of Energy, the National Aeronautics and Space
Administration, the Japanese Monbukagakusho, the Max Planck Society,
and the Higher Education Funding Council for England.  The SDSS Web
Site is http://www.sdss.org/.  The SDSS is managed by the
Astrophysical Research Consortium for the Participating
Institutions. The Participating Institutions are the American Museum
of Natural History, Astrophysical Institute Potsdam, University of
Basel, University of Cambridge, Case Western Reserve University,
University of Chicago, Drexel University, Fermilab, the Institute for
Advanced Study, the Japan Participation Group, Johns Hopkins
University, the Joint Institute for Nuclear Astrophysics, the Kavli
Institute for Particle Astrophysics and Cosmology, the Korean
Scientist Group, the Chinese Academy of Sciences (LAMOST), Los Alamos
National Laboratory, the Max-Planck-Institute for Astronomy (MPIA),
the Max-Planck-Institute for Astrophysics (MPA), New Mexico State
University, Ohio State University, University of Pittsburgh,
University of Portsmouth, Princeton University, the United States
Naval Observatory, and the University of Washington.

\begin{deluxetable}{lccc}
\tablewidth{0pt}
\tablecaption{Galaxy Completeness within 200 Mpc
\label{table:surveycompare}}
\tablehead{
\colhead{Survey Type}              &
\colhead{Proposed Depth}           &
\colhead{min($f_{\rm SFR}$)}       &
\colhead{min($f_{\rm M_{\star}}$)} 
}
\startdata
\Ha\ (ideal spectroscopic\tablenotemark{a})                  & $2\times 10^{-15}$ erg cm$^{-2}$ s$^{-1}$ & 97$\%$ & 58$\%$  \\
-                  & $1\times 10^{-15}$ erg cm$^{-2}$ s$^{-1}$ & 97$\%$ & 59$\%$  \\
\Ha\ (realistic imaging, optimistic\tablenotemark{b})  & $2\times 10^{-15}$ erg cm$^{-2}$ s$^{-1}$ & 95$\%$ & 47$\%$  \\
-  & $1\times 10^{-15}$ erg cm$^{-2}$ s$^{-1}$ & 96$\%$ & 52$\%$  \\
\Ha\ (realistic imaging, pessimistic\tablenotemark{c}) & $2\times 10^{-15}$ erg cm$^{-2}$ s$^{-1}$ & 76$\%$ & 33$\%$  \\
- & $1\times 10^{-15}$ erg cm$^{-2}$ s$^{-1}$ & 90$\%$ & 49$\%$  \\
\HI\                                           & $0.7\,{\rm mJy}\times 100\,{\rm kHz}$     & 93$\%$ & 44$\%$ \\
-                                           & $0.35\,{\rm mJy}\times 100\,{\rm kHz}$     & 95$\%$ & 49$\%$ 
\enddata

\tablecomments{We give the minimum completeness to SFR and stellar
mass based on our strawman \Ha\ and \HI\ surveys.  The \Ha\ flux limit
corresponds to a luminosity of $L_\Ha=10^{40}$ erg s$^{-1}$ at 200
Mpc.  The \HI\ flux limit corresponds to a luminosity of $L_{\rm
H\,I}=3\times 10^{31}$ erg s$^{-1}$, or an \HI\ mass of
$M_{\rm H\,I}\approx 10^9$ M$_\odot$.}
\tablenotetext{a}{This scenario assumes no Balmer absorption and that
all of the flux from extended sources is recovered (dotted line in Fig.~\ref{fig:Ha_complete}).}
\tablenotetext{b}{This scenario includes the effects of Balmer
absorption, but assumes that all of the flux from extended sources is
recovered (dashed line in Fig.~\ref{fig:Ha_complete}).}
\tablenotetext{c}{This scenario includes the effects of Balmer
absorption, but assumes that none of the flux from extended sources is
recovered (solid line in Fig.~\ref{fig:Ha_complete}).}
\end{deluxetable}

\end{document}